\documentstyle[12pt]{article}

\catcode`@=12
\begin{document}
\begin{titlepage}
\thispagestyle{empty}
\begin{center}
\mbox{}
\vskip 1.5in
{\Large \bf Neutrino Mass Matrices in Models with
Horizontal Symmetries \\}
\vspace{0.7in}
{\bf Asim K. Ray and Utpal Sarkar\\}
\bigskip
{\sl Department of Physics, Visva-Bharati University,
Santiniketan 731 235, India\\}
\bigskip
\end{center}

\begin{abstract}

We have studied the most general neutrino mass matrices in
models with SU(2) and SU(3) horizontal symmetries. Without
going into the details of the models it is possible to
write down the effective operators, which predict the
structure of the Majorana neutrino mass matrices.
Unlike other extensions of the standard model, the structure
is now independent of the effective Yukawa couplings and
depends entirely on the Higgs which gives mass to the other
fermions. In the
case of SU(3) symmetries the lowest dimensional operators
are forbidden requiring a low mass scale for lepton number
violation.
\end{abstract}
\end{titlepage}

The structure of fermion masses and mixing is not predicted
in the standard model because all generations are treated
similarly. All the fermion masses and mixing are free
parameters, which are supposed to be determined by
experiments. To eliminate this uncertainty and to understand
the generation structure several approaches has been
considered \cite{gen}. One such approach assumes that
the fermions of different generations are related by some
gauge symmetry group, called the horizontal symmetry
\cite{h}. The vacuum expectation
values ($vev$) of the Higgs would then
give us the structure of the fermion masses and mixing.

So far most of the searches for physics beyond the standard
model has met with negative results except for the neutrino
sector. The neutrino mass is predicted to be zero in the
standard model, but a solution to the atmospheric neutrino
anomaly \cite{atm} has evidenced a nonvanishing mass for the
neutrinos. This result has been supported by the solar
neutrino results \cite{sol}. Even the Laboratory experiments started
supporting this result predicting a neutrino mass and
mixing required to explain the observed neutrino
oscillations \cite{k2k,lsnd}. The developments in the neutrino physics
over the past few years have already established some
possible structures for the neutrino mass matrix.

Since the horizontal symmetric models are some of the possible
extensions of the standard model, it is most natural to
study the horizontal symmetries in the light of the recent
results from neutrino physics. In this article we point out
that it is possible to make some general comments about
the neutrino masses and mixing in some of the models
of gauged horizontal symmetries. We shall discuss the
nonabelian $SU(2)_H$, $SU(3)_H^V$ and
$SU(3)_H^{VL}$ horizontal gauge symmetries, although
our results could be generalized to other symmetries as
well.

We start with a brief introduction to the horizontal
symmetries, we are going to discuss. We shall include
the right handed neutrinos and their interactions while
studying the neutrino masses. In the $SU(2)_H$
horizontal symmetry all the left and right handed
fermions and antifermions transform as triplets.
The fermion bilinears
can thus be a singlet, tripplet or a 5-plet and hence the
standard model Higgs doublet fields required to give
fermion masses have to transform as a
singlet $\phi^{a1} [1,2,1,1]$ (trace), triplet
$ \phi^{a2}_k [1,2,1,3]$ (antisymmetric tensor) or a
5-plet $\phi^{a3}_{ij} [1,2,1,5]$ (traceless symmetric
tensor of rank 2) under $SU(2)_H$.
So, we need one or more of the
Higgs scalars $\phi^{a1}$, $\phi^{a2}_i$ or $\phi^{a3}_{ij}$
to write down the Yukawa couplings
for the fermion masses and mixing
\begin{eqnarray}
{\cal L}_Y &=& f^u_{1} \delta_{ij}
\bar q_{iL} u_{jR} {\phi^{a1}}^\dagger + \epsilon_{ijk}
f^u_{3} \bar q_{iL} u_{jR} {\phi^{a2}_k}^\dagger +
f^u_{5} \bar q_{iL} u_{jR} {\phi^{a3}_{ij}}^\dagger  \nonumber \\
&&+ f^d_{1} \delta_{ij}
\bar q_{iL} d_{jR} \phi^{a1} + \epsilon_{ijk}
f^d_{3} \bar q_{iL} d_{jR} \phi^{a2}_k +
f^d_{5} \bar q_{iL} d_{jR} \phi^{a3}_{ij}  \nonumber \\
&&+ f^e_{1} \delta_{ij}
\bar \ell_{iL} e_{jR} \phi^{a1} + \epsilon_{ijk}
f^e_{3} \bar \ell_{iL} e_{jR} \phi^{a2}_k +
f^e_{5} \bar \ell_{iL} e_{jR} \phi^{a3}_{ij} + h.c.
\end{eqnarray}
Here $i,j,k=1,2,3$ are $SU(2)_H$ indices. The $vev$s of the
different components of a Higgs scalar now determine the
structure of the fermion mass matrices. The $SU(2)_H$
symmetry now requires same Yukawa couplings for all
generations. The singlet
$\phi^{a1} \delta_{ij}$ being the trace of the matrix in the generation
space, a $vev$ of the field $\phi^{a1}$
can only give equal masses to all the fermion
generations with no mixing. Thus we need at least two
or more of the Higgs fields $\phi^{a2}_k$ and
$\phi^{a3}_{ij}$. In addition to these standard model Higgs doublets
there are some singlets fields which breaks the horizontal
group. Their inclusion do not change our analysis.

There are two possible realizations of the $SU(3)$ horizontal
symmetries. In the vector-like case the left handed
fermions and antifermions transform as triplets, while the
right handed fermions and antifermions transform as
antitriplets of $SU(3)_H^{VL}$. In this case we need to
introduce mirror fermions to make the theory anomaly free
and hence consistent.
For every left handed field there is a right handed mirror
field which has the same transformation under all gauge
symmetries. In some theories such mirror fermions
come out naturally and has interesting consequences, but
otherwise this doubles the number of fermions. The mirror
fermions are decoupled by some discrete symmetry
from interacting with usual fermions
and hence escape all detections. We shall not discuss the
mirror sector here.

To contribute to the fermion masses
the standard model Higgs doublet now could be an antitriplet
$\phi^{b1}_{[ij]} [1,2,1,\bar 3]$ (antisymmetric rank 2 tensor)
or a sextet $\phi^{b2}_{\{ij\}} [1,2,1,6]$ (symmetric rank 2
tensor) under $SU(3)_H^{VL}$. To give masses to both the up
and down quark sector we need another
triplet $\tilde \phi^{b1}_{[ij]} [1,2,1,3]$ and antisextet
$\tilde \phi^{b2}_{[ij]} [1,2,1,\bar 6]$ of $SU(3)_H^{VL}$,
which carry same hypercharge as $\phi^{b1}_{\{ij\}}$
and $\phi^{b2}_{[ij]}$ respectively.
The Yukawa couplings are now given by
\begin{eqnarray}
{\cal L}_Y &=& h^u_3 \bar q_{iL} u_{jR}
{\tilde \phi^{b1^\dagger}_{[ij]}}
+ h^d_3 \bar q_{iL} d_{jR} \phi^{b1}_{[ij]} +
h^e_3 \bar \ell_{iL} e_{jR} \phi^{b1}_{[ij]}  \nonumber \\
&&+ h^u_6 \bar q_{iL} u_{jR} {\tilde \phi^{b2^\dagger}_{\{ij\}}} +
h^d_6 \bar q_{iL} d_{jR} \phi^{b2}_{\{ij\}}
+ h^e_6 \bar \ell_{iL} e_{jR} \phi^{b2}_{\{ij\}} +h.c.
\end{eqnarray}
Since $\phi^{b1}_{[ij]}$ [$\tilde \phi^{b1}_{[ij]}$] and
$\phi^{b2}_{\{ij\}}$
[$\tilde \phi^{b2}_{\{ij\}}$] are most general anti-symmetric and
symmetric matrices, they can provide the required fermion
mass matrices and their mixings.

The second realization of $SU(3)$ is different, because it
now requires a right handed neutrino to cancel anomaly.
In the $SU(3)_H^V$ the fermions are vectorial.
Both the left and right handed fermions transform as triplets
under $SU(3)_H^V$ and all anti-fermions transform as
anti-triplets. This implies that the standard model Higgs doublet
that gives mass to the fermions after symmetry breaking could
be a singlet $\phi^{c1} [1,2,1,1]$ or an octet
$\phi^{c2}_{ij} [1,2,1,8]$ under $SU(3)_H^V$,
\begin{eqnarray}
{\cal L}_Y &=& g^u_{1} \delta_{ij} \bar q_{iL} u_{jR} {\phi^{c1}}^\dagger
+ g^d_{1} \delta_{ij} \bar q_{iL} d_{jR} \phi^{c1} +
g^e_{1} \delta_{ij} \bar \ell_{iL} e_{jR} \phi^{c1}  \nonumber \\
&&+ g^u_8 \bar q_{iL} u_{jR} {\phi^{c2}_{ij}}^\dagger
+ g^d_8 \bar q_{iL} d_{jR} \phi^{c2}_{ij} +
g^e_8 \bar \ell_{iL} e_{jR} \phi^{c2}_{ij} + h.c.
\end{eqnarray}
The singlet Higgs $\phi^{c1}$ gives equal masses to all generations
with no mixing, but together with the octet Higgs they can
produce all the required masses and mixing for the fermions.
In the minimal model there are no new doublet Higgs scalar
in addition to the $\phi^{c1}$ and
$\phi^{c2}$. To break the $SU(3)_H^V$ horizontal
symmetry some standard models singlet Higgs scalars are required
which shall be discussed later.

The main features of all the
horizontal symmetries is that the Higgs doublets are required
to be such that the generation structure comes only from the $vev$s
of the doublet Higgs. The Yukawa couplings do not carry any generation
index. We shall restrict ourselves to Higgs doublets which are
required by the minimal models as discussed above.

We now turn to the question of neutrino masses in these
horizontal symmetric models. The natural scenario of small neutrino
masses is to have Majorana masses which are different
realizations of a single effective operator \cite{path},
\begin{equation}
{\cal L}_{eff} = y^{eff}_{ij} \ell_{iL} \ell_{jL} \phi \phi  .
\end{equation}
where $\phi$ is the usual Higgs doublet.
The effective coupling $y^{eff}_{ij}$ of this interaction includes
the lepton number violating large scale $M$ in the denominator.
If we write $y^{eff} \sim f/M$, then for $f \sim 10^{-3}$ we get
$M \sim 10^{12}$ GeV to explain the neutrino masses required by
the atmospheric neutrino anomaly. For radiative models $f$ would have
a smaller value and hence $M$ could be lower.
The effective coupling $y^{eff}$ will
depend on the actual realization of this operator. The
structure of the neutrino mass matrix
\begin{equation}
{\cal M}_{\nu ij} = y^{eff}_{ij} <\phi>^2
\end{equation}
depends completely on the effective Yukawa coupling.

The main point in our paper is that in theories with horizontal
symmetries, the structure of the effective operators remain same,
but the effective coupling is now completely independent of the
generation structure. All the information about the generation
structure are contained in the structure of the Higgs doublets.
In other words, the structure of the
neutrino mass matrix can be determined
completely if we know how the Higgs doublets transform
under the horizontal symmetric group. The only assumption
we are making here is that for phenomenological reasons the
horizontal symmetry is assumed to be broken at a scale
much lower than the lepton number violating scale $M$.
We assume that the mass of the gauge bosons of the horizontal
symmetry is about 100 TeV, to have consistency with the
lower bound on their mass coming from processes like
$K_L \to \mu^\pm + e^\mp$.
If the horizontal symmetry is broken at a
very high scale when the lepton number symmetry is also broken
or higher, our result does not hold. Although we shall be
discussing only few horizontal symmetry groups, but the
above observation is true for any other horizontal symmetry
groups including the abelian symmetries.

To understand this point, let us consider the example of see-saw
mechanism \cite{seesaw}. In general the heavy neutrino mass matrix could
have nontrivial structure in the generation space.
But in case of horizontal
symmetric models, at the lepton number violating scale the
horizontal symmetries are still exact. So,
the horizontal symmetries will prevent
nontrivial structures of the heavy neutrino mass matrix. Only
diagonal mass matrices are allowed for the right handed
neutrinos.

In the minimal version of $SU(2)_H$, there is no right handed
neutrino. One may add to the theory right handed neutrinos
$N_{iR}$, which are singlets or even triplets under $SU(2)_H$.
When the right handed neutrinos are three singlets, one
may add a general Majorana mass term with nonvanishing
off-diagonal terms. However, since $N_{iR}$ does not have
any gauge interactions, without loss of generality we can
diagonalise the mass matrix and work in a basis in which
the heavy right handed neutrino mass matrix is diagonal.

In $SU(2)_H$ there are three standard model
doublet Higgs scalars, a singlet $\phi^{a1}$, a triplet
$\phi^{a2}_i$ and a 5-plet $\phi^{a3}_{ij}$. So for the
singlet right handed neutrinos the Dirac
neutrino masses could come only from the triplet Higgs
scalar
\begin{equation}
{\cal L}_D = f^\nu_j \bar \ell_{iL} N_{jR}
{\phi^{a2}_i}^\dagger .
\end{equation}
If neutrino mass is generated through see-saw mechanism,
the left handed neutrino mass matrix would come only
from the triplet Higgs. We shall not restrict our discussions
of neutrino masses to only one mechanism and hence we
shall present a more  general analysis.

The right handed neutrinos could also be triplets.
We assumed that when the right handed neutrinos
get mass at a scale $M$ breaking lepton number,
the horizontal symmetry is exact. This implies that
the right handed neutrino mass matrix comes from a
singlet Higgs and hence the mass matrix is diagonal.
Hence the low energy neutrino mass matrix
again depends only on the standard model Higgs scalars
appearing in the effective Majorana mass term and not on
the effective Yukawa coupling constant $y^{eff}$. The Dirac neutrino
mass term will now have contributions from all the Higgs
scalars
\begin{equation}
{\cal L}_D = f_1^\nu \delta_{ij} \bar \ell_{iL} N_{jR}
{\phi^{a}_i}^\dagger
+ f_3^\nu \epsilon_{ijk} \bar \ell_{iL} N_{jR}
{\phi^{a2}_k}^\dagger
+ f_5^\nu \bar \ell_{iL} N_{jR}
{\phi^{a3}_{ij}}^\dagger .
\end{equation}
Thus $vev$s of all the Higgs scalars would
contribute to the left handed Majorana neutrino mass
matrix.

We can now come back to the effective operator contributing
to the left handed neutrino Majorana mass matrix. These
operators do not depend on the existence of the right handed
neutrinos.
Since the left handed neutrinos are triplets under $SU(2)_H$
the Majorana mass term $\ell_{iL} \ell_{jL}$ can be a singlet,
triplet or a 5-plet. A singlet contribution can come from
an operator $y^{eff} \ell_{iL} \ell_{jL} \phi^{a1} \phi^{a1}$,
but since this can give only a diagonal neutrino
mass it is not sufficient. The triplet combination
$y^{eff} \epsilon_{ijk} \ell_{iL} \ell_{jL} \phi^{a1} \phi^{a12}_k$
vanishes identically since the neutrino mass matrix is symmetric.
Thus only the 5-plet combination is important. In the following
we list all possible operators which can contribute to the 5-plet
combination of the neutrino mass matrix
\begin{eqnarray}
{\cal L}^1_{eff} &=&
y^{eff} \ell_{iL} \ell_{jL} \phi^{a2}_i \phi^{a2}_j
\nonumber \\
{\cal L}^2_{eff} &=&
y^{eff} \ell_{iL} \ell_{jL} \phi^{a1} \phi^{a3}_{ij}
\nonumber \\
{\cal L}^3_{eff} &=&
y^{eff} \ell_{iL} \ell_{jL} \phi^{a3}_{ik} \phi^{a3}_{jk}  .
\end{eqnarray}
An 5-plet of $SU(2)_H$ has isospin quantum number 2. Hence
it has 5 projections, which are the 5-components. We can give
$vev$s to any one of these components and hence there could
be only 5 parameters that can determine the structure of the
neutrino mass matrix. This is given by the matrix
\begin{equation}
\pmatrix{a & b & c \cr b & {c \over \sqrt{2}} & d \cr c & d & e} .
\end{equation}
The fact that the (22) component is same as the (13) or (31)
component (modulo a factor of $\sqrt{2}$) restricts the possible
textures of the neutrino mass matrix. Otherwise the mass
matrix is very general and can explain the present data from
atmospheric neutrino, solar neutrino and laboratory experiments
for suitable value of the parameters coming from the
$vev$s of the Higgs scalars \cite{nu}. The
simplest choice would be when only the singlet give the charged
lepton mass and hence diagonal. Otherwise we first have to
diagonalise the charged lepton mass matrix to get the actual
neutrino mass matrix.

For $SU(3)_H^{VL}$ one can only add singlet right handed
neutrinos so as to give them Majorana mass with only
$SU(3)_H^{VL}$ singlet Higgs scalars. Since these neutrinos
do not have any gauge interaction, the mass matrix can be
diagonalised without loss of generality. The Dirac mass of
the neutrinos can then come from both the triplet
$\tilde \phi^{b1}_{[ij]} $ and anti-sextet
$\tilde \phi^{b2}_{\{ij\}} $ Higgs scalars
\begin{equation}
{\cal L}_D=h_3^\nu \bar \ell_{iL} N_{jR}
{\tilde \phi^{b1^\dagger}_{[ij]}}
+ h^u_6 \bar \ell_{iL} N_{jR} {\tilde \phi^{b2^\dagger}_{\{ij\}}}  .
\end{equation}
The effective operator will not depend on the existence of
the right handed neutrinos.

We now study the effective operators contributing to the
Majorana neutrino masses for $SU(3)_{VL}$.
Since the left handed neutrinos are triplets and the mass term
is symmetric, only a sextet combination of the Higgs scalars
would contribute to the effective operator. Possible operators
are
\begin{eqnarray}
{\cal L}^1_{eff} &=& y^{eff} \epsilon_{ikl} \epsilon_{jmn}
\ell_{iL} \ell_{jL}
\phi^{b1}_{[kl]} \phi^{b1}_{[mn]} \nonumber \\
{\cal L}^1_{eff} &=& y^{eff} \epsilon_{ikm} \epsilon_{jln}
\ell_{iL} \ell_{jL} \phi^{b2}_{\{kl\}} \phi^{b2}_{\{mn\}}
\end{eqnarray}
Since the effective mass term is a sextet, it is given by 6
parameters in the matrix
\begin{equation}
\pmatrix{a & b & c \cr b & d & e \cr c & e & f} .
\end{equation}
It is thus clear that this mass matrix is the most general
symmetric mass matrix and hence can explain any
experiment with proper choice of these six parameters,
which are given by the $vev$s of the Higgs doublets.

The $SU(3)_H^V$ case is most interesting. As in the
$SU(3)_H^{VL}$ case, the right handed neutrinos can
only be singlets, otherwise they cannot get Majorana
mass before $SU(3)_H^V$ is broken. The left handed
leptons are again triplet, so the Dirac mass term would
require a triplet Higgs scalar, which is not there in the
model. Hence the right handed neutrinos decouple from
the left handed neutrino mass matrix. We shall come
back to this point later.

Since the left handed neutrinos are triplets, the effective
mass term is a sextet as in the previous case. But now
the Higgs scalars in the minimal model are singlet and
octet. Hence there are no dimension five operators, which
can allow neutrino masses. If we now include a Higgs,
which is standard model singlet and triplet
under $SU(3)_H^V$ $\eta_{\{ij\}} [1,1,0,\bar 6]$,
which can break the horizontal symmetry group,
then we can have the effective operators
\begin{eqnarray}
{\cal L}^1_{eff} &=& y^{eff} \ell_{iL} \ell_{jL}
\phi^{c1} \phi^{c1} \eta_{\{ij\}} \nonumber \\
{\cal L}^2_{eff} &=& y^{eff} \ell_{iL} \ell_{jL}
\phi^{c2}_{kl} \phi^{c2}_{il} \eta_{\{kj\}} \nonumber \\
{\cal L}^3_{eff} &=& y^{eff} \ell_{iL} \ell_{jL}
\phi^{c1} \phi^{c2}_{ik} \eta_{\{kj\}}  .
\end{eqnarray}
Now the operators are of dimension 6 and hence
suppressed by one extra power of the lepton number
violating scale $M$. In case of dimension 5 operators the
lepton number violating scale was $M \sim 10^{12}$ GeV.
But now the required lepton number violating scale
would be about $M \sim 10^8$ GeV. Since we cannot
construct a dimension 5 operator in this case, it is
obvious the mechanism for neutrino mass has to be
somewhat different from the canonical case. This is
clear from the fact that there is no Dirac mass term
now (which will immediately forbid the see-saw
mechanism). For example, one can now generate a
neutrino mass in the triplet Higgs mechanism \cite{trip}, where
the triplet Higgs is a sextet under $SU(3)_H^V$.
This result will change if one
considers nonminimal model by including a sextet
and triplet standard model doublet Higgs scalars.

The lepton number violating scale could be further
lowered, if the horizontal symmetry is broken by a
triplet Higgs $\omega_i [1,1,0,\bar 3]$ instead of a
sextet Higgs. In this case the effective operators are
\begin{eqnarray}
{\cal L}^1_{eff} &=& y^{eff} \ell_{iL} \ell_{jL}
\phi^{c1} \phi^{c1} \omega_i \omega_j  \nonumber \\
{\cal L}^2_{eff} &=& y^{eff} \ell_{iL} \ell_{jL}
\phi^{c2}_{kl} \phi^{c2}_{il}  \omega_k \omega_j \nonumber \\
{\cal L}^3_{eff} &=& y^{eff} \ell_{iL} \ell_{jL}
\phi^{c1} \phi^{c2}_{ik}  \omega_k \omega_j  .
\end{eqnarray}
These operators have dimension 7 and hence they
are suppressed by third power of lepton number violating
scale. This implies that the lepton number violating scale
$M$ could be as low as $M \sim 1000$ TeV.
In this case the left handed
neutrino mass would be determined by the sextet structure
of the effective Higgs and hence the mass matrix would
appear to be similar to $SU(3)_H^{VL}$ model.

In summary, we pointed out that in models with gauged
horizontal symmetries it is possible to get the structure
of the neutrino mass matrix without knowing the
mechanism which generates the neutrino mass. This is
because the effective operator for the neutrino masses
contains all the generation indices in the Higgs $vev$s
and the effective Yukawa coupling is independent of the
generation structure. We studied the neutrino mass matrices
in the $SU(2)_H$ and $, SU(3)_H^V$ and $SU(3)_H^{VL}$
horizontal symmetric models.

\end{document}